\documentclass[a4paper,11pt]{article}
\pdfoutput=1 

\usepackage{jheppub} 

\usepackage[T1]{fontenc} 

\title{\boldmath Generating Entangled Inflationary Quantum States}

\author{R. Holman,}
\author{Benoit J. Richard}
\affiliation{Minerva Schools at KGI,\\1145 Market Street, San Francisco, CA 94103, USA}

\emailAdd{rholman@minerva.kgi.edu}
\emailAdd{brichard@minerva.kgi.edu}

\abstract{Entangled inflationary quantum states offer an interesting alternative to the standard Bunch-Davies vacuum. While they can be viewed from the point of view of a theory of effective initial states, it has been unclear exactly how such states might be generated. Using a model involving two scalar fields $\Phi,\Sigma$ we show that if {\em both} fields have non-zero time dependent expectation values $\phi,\sigma$ respectively, and if the interaction potential $V(\Phi,\Sigma)$ when evaluated at these expectation values has $\partial_{\phi}\partial_{\sigma}V(\phi, \sigma)\neq 0$, this can serve as a source term for the entanglement kernel describing the state within the Gaussian approximation around the expectation values. We also show how cubic interactions between the metric perturbation $\zeta$ and a scalar $\Sigma$ can be used to generate a source term for entanglement.}

\begin{document} 
\maketitle
\flushbottom

\section{Introduction}
\label{sec:intro}

One of the most important realizations in modern cosmology was that the statistics of temperature fluctuations in the cosmic microwave background (CMB) provide us a Rosetta Stone of a sort, allowing us to translate late-time observations into statements about the very early universe. Within the context of the inflationary universe paradigm, CMB data becomes exceedingly useful, affording us the opportunity to at least partially distinguish between viable models of inflation\cite{Ade:2015lrj,Akrami:2018odb}, as well as pointing out interesting potential anomalies\cite{Ade:2015hxq} such as the large scale hemispheric anomaly. Attempts to explain such anomalies may be able to further narrow down the space of allowed possible inflationary models, or perhaps falsify the paradigm entirely. 

One important ingredient that factors into the connection between inflation and correlation functions of CMB temperature fluctuations is the quantum state of the scalar and metric tensor fluctuations $\zeta$ and $h_{i j}$, respectively. From these quantum states, quantum correlation functions can be calculated and then, due to the decohering nature of the inflationary expansion, these fluctuations then become classical. Within the context of inflation then, CMB fluctuations as well as large scale structure represent quantum mechanics writ on the largest measured scales!

The standard lore concerning the quantum state of metric perturbations is that in the far past the relevant mode functions, corresponding to physical wave numbers deep within the inflationary horizon, should behave as if they were in flat space. The solution to the mode equations that matches this boundary condition is the so-called Bunch-Davies (BD) state\cite{Bunch:1978yq}; this is an infinite order adiabatic vacuum\cite{Birrell:1982ix} which makes them the best approximation, in a curved spacetime,  to a state devoid of particles.

There are, however, valid reasons to explore the possibility that states other than the Bunch-Davies one might be relevant during inflation\cite{Albrecht:2018hoh}. Deviations from the BD state generally require that the universe has been in a near de Sitter state for at most the 60 e-folds that we have information about. More than this and the energy density in the state (which can sometimes be thought of as coming from a distribution of particles created from the BD vacuum) will be large enough to prevent inflation from occurring. It is worth emphasizing that {\em none} of the available cosmological data requires inflation to last more than the requisite 60 e-folds needed to solve the flatness and horizon problems. There are naturalness arguments for a larger number of e-folds\cite{Kaloper:2018zgi}, but these we feel are more aesthetic than experimental in nature. In this work we will restrict inflation to the bare minimum number of e-folds in order to entertain the possibility of non-BD inflationary states. 

Recently a new category of inflationary states has been discussed, the so-called {\em entangled} states\cite{Albrecht:2014aga}. These are premised on the existence of scalar fields other than the inflaton, such as the standard model Higgs as an example, whose quantum states can become entangled with that of the inflaton. The effects of entanglement on the power spectrum\cite{Albrecht:2014aga}, the bispectrum\cite{usbispectrum} have been examined, as has the possibility of entangling the scalar and tensor fluctuations\cite{Bolis:2016vas,Collins:2016ahj} in an attempt to explain some of the large scale anomalies. In these works the origin of these states was not specified; rather, they were treated more from the point of view of an effective theory of states, where we would allow all possible ``operators'' consistent with given symmetries in the state. Our goal in this work is to show how interactions between the fields can induce entanglement starting from uncorrelated states for each field (see refs.\cite{Bolis:2018jmo,Phillips:2014yma,Kanno:2014ifa} for some other approaches to this problem). Since the states described in ref.\cite{Albrecht:2014aga} (hereafter known as ABH states) are Gaussian the question arises as to how to consistently include the effect of interactions. We do this by extending the form of the state to include the effect of a non-vanishing, time-dependent expectation value for both fields. Doing this then allows us to consider a state describing both the evolution of these expectation values, as well as of the fluctuations around them; it is this latter description that has a Gaussian approximation associated with it.  We find that doing this induces a source term for the entanglement kernel which appears when the state is described within the Schr\"odinger picture so that even when we start off with zero entanglement, interactions will generate it. 

The next section deals with how entanglement might be generated by interactions between two scalar fields. Section \ref{sec:entzeta} then turns to the possibility that the metric perturbation $\zeta$ could be entangled with a scalar field and also discusses how $\zeta$ and the tensor fluctuations $h_{i j}$ might find themselves entangled. We then conclude.

\section{Schr\"odinger Picture Field Theory Description of ABH states}
\label{sec:SpicABH}

In this section we both review and extend the description of entangled field theoretic states now allowing for time-dependent expectation values for both fields. We also use a density matrix formalism, as in ref.\cite{Boyanovsky:1993xf}; though we still consider pure states, we could use this to allow for more general mixed states if need be. From our perspective in this work, the density matrix formalism allows for cancellation of terms that would otherwise have to be dealt with by a rephasing of the wavefunctional. 

We begin by considering two scalar fields $\Phi,\Sigma$ evolving in a background FRW geometry with scale factor $a(\eta)$, where $\eta$ is conformal time. We set $C(\eta) = a^2(\eta)$ and write down an action where both scalars have canonical kinetic terms:

\begin{eqnarray}
\label{eq:action}
S =  \int d^4 x\ C^2(\eta) & & \left\{\frac{1}{2 C(\eta) }  \left(\Phi'(\vec{x},\eta)^2-(\nabla \Phi(\vec{x}, \eta))^2\right)+\right . \nonumber\\
 & &  \left .  \frac{1}{2 C(\eta) } \left(\Sigma'(\vec{x},\eta)^2-(\nabla \Sigma(\vec{x}, \eta))^2\right)-V(\Phi,\Sigma) \right\}.
\end{eqnarray}
Here primes denote conformal time derivatives. 

We should note that since as we will see below, the state will be taken to be Gaussian, consistency will require that we keep at most the quadratic order terms in the fluctuations $\gamma(\vec{x},\eta)$, $\chi(\vec{x},\eta)$ about $\langle \Phi(\vec{x}) \rangle(\eta)\equiv \phi(\eta)$ and $\langle \Sigma(\vec{x})\rangle(\eta)\equiv\sigma(\eta)$, respectively. Ultimately, the action will be quadratic in $\gamma(\vec{x},\eta), \chi(\vec{x}, \eta)$, but with mass terms that depend on $\phi$ and $\sigma$ in general. 

Starting from the action in eq. \eqref{eq:action} we can obtain the momenta canonically conjugate to $\Phi$ and $\Sigma$, $\Pi_{\Phi},\ \Pi_{\Sigma}$ respectively, and use these to construct the Hamiltonian of the system:
\begin{subequations}\label{eq:momHam}
\begin{align}
\label{eq:momHam:momenta}
& \Pi_{\Phi} (\vec{x},\eta)  = C(\eta) \Phi'(\vec{x},\eta),\  \Pi_{\Sigma} (\vec{x},\eta) = C(\eta) \Sigma'(\vec{x},\eta),
\\
\label{eq:momHam:hamiltonian}
& H  = \int d^3 x\ \left\{\frac{\Pi_{\Phi}^2}{2 C(\eta)}+\frac{\Pi_{\Sigma}^2}{2 C(\eta)}+\frac{C(\eta)}{2}\left(\nabla \Phi\right)^2+\frac{C(\eta)}{2}\left(\nabla \Sigma\right)^2+C^2 (\eta)V(\Phi,\Sigma)\right\}.\\ \nonumber
\end{align}
\end{subequations}

As mentioned above the entangled state we will use is best described within the Schr\"odinger functional formalism\cite{Boyanovsky:1993xf}. We envision a density matrix $\rho(\eta)$ with matrix elements taken on field configurations $\Phi(\cdot),\Sigma(\cdot)$ and $\tilde{\Phi}(\cdot),\tilde{\Sigma}(\cdot)$ specified at conformal time $\eta$. This is used to construct the probability distributions for finding the given field configurations as well as all relevant expectation values of field and momentum operators. 

\subsection{Rolling Expectation Values and their Equations of Motion}
\label{sec:SpicABH:rolling}

Before setting down the explicit form of the density matrix describing the ABH state we will use, we need to allow for the possibility of a time evolving or rolling expectation value for the fields we are entangling. In this section, we keep the formulation rather general and construct these expectation values and their equations of motion. At the end, we truncate the system to keep only terms that are quadratic in the fluctuations of the fields about these expectation values.

Recall that in the Schr\"odinger picture, operators such as the field operators $\Phi(\vec{x}),\Sigma(\vec{x})$ are time independent; all time dependence resides in the state described by $\rho(\eta)$. The time dependent expectation values of $\Phi(\vec{x})$ and $\Sigma(\vec{x})$ are then given by:

\begin{equation}
\label{eq:expectval}
\phi(\eta)\equiv {\rm Tr}\left(\rho(\eta) \Phi(\vec{x})\right),\ \sigma(\eta)\equiv {\rm Tr}\left(\rho(\eta) \Sigma(\vec{x})\right).
\end{equation}
Let us focus our discussion on $\phi(\eta)$; everything we do will have a direct analog for $\sigma(\eta)$.

The fact that $\phi$ only depends on $\eta$ and not on $\vec{x}$ as well is due to the choice of a spatially translationally invariant state $\rho$. If we quantized the system in a spatial box, we would have performed a spatial averaging over the box volume to construct $\phi$.  

The Liouville equation $i\partial_{\eta} \rho(\eta) = \left[H,\rho(\eta)\right]$, where $H$ is the Hamiltonian operator in \eqref{eq:momHam:hamiltonian} then allows us to generate the equation of motion for $\phi$:

\begin{subequations}\label{eq:eqsom}
\begin{align}
\label{eq:eqsom:momenta}
\partial_{\eta}\phi(\eta) & =-i  {\rm Tr}\left(i\partial_{\eta} \rho(\eta) \Phi(\vec{x})\right)=-i{\rm Tr}\left(\left[H,\rho(\eta)\right]\Phi(\vec{x})\right)\nonumber \\
& = -i{\rm Tr}\left(\rho(\eta) \left[\Phi(\vec{x}),H\right]\right)={\rm Tr}\left(\rho(\eta) \frac{\Pi_{\Phi}(\vec{x}) }{C(\eta)}\right)\equiv \frac{\pi_{\Phi}(\eta)}{C(\eta)},
\\
\label{eq:eqsom:etaderivmom}
\partial_{\eta}\pi_{\Phi}(\eta) & =-{\rm Tr}\left(i\partial_{\eta} \rho(\eta) \Pi_{\Phi}(\vec{x})\right)=-i {\rm Tr}\left(\left[H,\rho(\eta)\right] \Pi_{\Phi}(\vec{x})\right)\nonumber \\
& = -i {\rm Tr}\left(\rho(\eta) \left[ \Pi_{\Phi}(\vec{x}),H\right]\right)=-C^2(\eta)  {\rm Tr}\left(\rho(\eta) \frac{\delta V(\Phi,\Sigma)}{\delta \Phi}\right).
\end{align}
\end{subequations}
Putting eqs.\eqref{eq:eqsom} together then yields the required equation of motion:

\begin{equation}
\label{eq:zeromodeeom}
\phi^{\prime \prime}(\eta) + D(\eta) \phi^{\prime}(\eta) + C(\eta) \left\langle\frac{\delta V(\Phi,\Sigma)}{\delta \Phi}\right\rangle=0,
\end{equation}
where $D(\eta) \equiv C'(\eta)\slash C(\eta)$, with a similar equation for $\sigma(\eta)$. 

As it stands, it's hard to do much more with eq.\eqref{eq:zeromodeeom}, since the expectation value will be difficult to evaluate in general. To go further, we need to somehow replace $\Phi$ and $\Sigma$ by $\phi$ and $\sigma$ respectively inside the potential term in the equations of motion. To do this we expand $\Phi, \Sigma$ about $\phi, \sigma$ in terms of fluctuation fields $\gamma(\vec{x}, \eta),\chi(\vec{x},\eta)$ respectively:

\begin{eqnarray}
\label{eq:fluctexp}
&\Phi(\vec{x}) = \phi(\eta)+\gamma(\vec{x}, \eta),\nonumber\\
&\Sigma(\vec{x}) = \sigma(\eta)+\chi(\vec{x}, \eta).
\end{eqnarray}
This definition implies that $\langle \gamma(\vec{x}, \eta)\rangle= \langle \chi(\vec{x}, \eta)\rangle=0$.

There's another point worth noting here: as mentioned above, $\Phi(\vec{x}),\Sigma(\vec{x})$ are Schr\"odinger picture operators and hence time independent. This means that the Schr\"odinger picture fluctuation fields $\gamma(\vec{x}, \eta),\chi(\vec{x},\eta)$ {\em must} have explicit time dependence in order to compensate for the time dependence in the backgrounds $\phi(\eta), \sigma(\eta)$. We can infer this time dependence by taking time derivatives of both sides of eq. \eqref{eq:fluctexp}:

\begin{equation}
\label{eq:timedervlfucts}
0=\phi^{\prime}(\eta)+\gamma^{\prime}(\vec{x},\eta)\Rightarrow \gamma^{\prime}(\vec{x},\eta)=-\phi^{\prime}(\eta),
\end{equation}
and likewise $\chi^{\prime}(\vec{x},\eta)=-\sigma^{\prime}(\eta)$.

Going back to the zero mode equation of motion eq.\eqref{eq:zeromodeeom}, we can now expand the last term around the point: $\left(\Phi, \Sigma\right)=(\phi, \sigma)$, and as mentioned before, we will only keep the quadratic terms in this expansion. Doing this gives us:

\begin{equation}
\label{eq:quadzeromodeeom}
\begin{split}
\phi^{\prime \prime}(\eta) + D(\eta) \phi^{\prime}(\eta) + C(\eta) \Bigg(\frac{\partial V(\phi, \sigma)}{\partial \phi} 
& \left. + \frac{1}{2} \frac{\partial^3 V(\phi, \sigma)}{\partial \phi^3}\langle \gamma^2\rangle \right. \\
& \left. + \frac{1}{2} \frac{\partial^3 V(\phi, \sigma)}{\partial \sigma^2 \partial \phi}\langle \chi^2\rangle+\frac{\partial^3 V(\phi, \sigma)}{\partial \sigma \partial \phi^2}\langle \chi \gamma\rangle\Bigg) =0, \right. \\
\sigma^{\prime \prime}(\eta) + D(\eta) \sigma^{\prime}(\eta) + C(\eta) \Bigg(\frac{\partial V(\phi, \sigma)}{\partial \sigma}
& \left. +\frac{1}{2} \frac{\partial^3 V(\phi, \sigma)}{\partial \sigma^3}\langle \chi^2\rangle \right. \\
&\left. + \frac{1}{2} \frac{\partial^3 V(\phi, \sigma)}{\partial \sigma \partial \phi^2}\langle \gamma^2\rangle+\frac{\partial^3 V(\phi, \sigma)}{\partial \sigma^2 \partial \phi}\langle \chi \gamma\rangle\Bigg) =0. \right.
\end{split}
\end{equation}
Note that we allow for the possibility of cross-correlators as $\langle \gamma \chi\rangle$ such as will occur if the state contains entanglement.

In order to solve these equations of motion, we need to specify the state $\rho(\eta)$ describing the behavior of the fluctuations and solve its equation of motion together with the ones above self-consistently. We turn to the description of the state next. 

\subsection{The Entangled State}
\label{sec:SpicABH:state}

It is sufficient for our purposes to work with a pure, though entangled state, so that we can view the density matrix described by the state as $\rho(\eta)=\left | \Psi(\eta)\rangle \langle \Psi(\eta)\right |$ for some state $\left | \Psi(\eta)\rangle \right .$ satisfying the Schr\"odinger equation. Doing this not only allows for the possibility of treating a mixed state\cite{RHBRmixedentanglement}, such as thermal one, but also simplifies some aspects of the calculation of the Liouville equation of motion. We can then  write the matrix elements of our putative state as:
\begin{equation}
\label{eq:densmatmatrixelts}
\langle \Phi, \Sigma \left | \rho(\eta) \right | \tilde{\Phi}, \tilde{\Sigma}\rangle = N(\eta)\exp\left(-\frac{1}{2}Q\left[\gamma, \chi; \tilde{\gamma}, \tilde{\chi}\right]+i P\left[\gamma, \chi; \tilde{\gamma}, \tilde{\chi}\right]\right),
\end{equation}
where $N(\eta)$ serves to normalize the state and we define $Q\left[\gamma, \chi; \tilde{\gamma}, \tilde{\chi}\right], P\left[\gamma, \chi; \tilde{\gamma}, \tilde{\chi}\right]$ as follows:
\begin{subequations}\label{eq:kernels}
\begin{align}
\label{eq:kernels:quadratic}
&Q\left[\gamma, \chi; \tilde{\gamma}, \tilde{\chi}\right] \equiv \int \frac{d^3 k}{(2\pi)^3}\ \left\{\left(\begin{array}{cc}\gamma_{\vec{k}},\ \chi_{\vec{k}}\end{array}\right)\hat{Q}\left(\begin{array}{c}\gamma_{-\vec{k}} \\\chi_{-\vec{k}}\end{array}\right)+\left(\gamma_{\vec{k}} \leftrightarrow \tilde{\gamma}_{\vec{k}} , \chi_{\vec{k}}\leftrightarrow \tilde{ \chi}_{\vec{k}}, \hat{Q}\leftrightarrow \hat{Q}^*\right)\right\} , \nonumber \\
\\ 
\label{eq:kernels:momentum}
& P\left[\gamma, \chi; \tilde{\gamma}, \tilde{\chi}\right] \equiv \int \frac{d^3 k}{(2\pi)^3}\ \left\{p_{\vec{k}\Phi}(\eta)\left(\gamma_{-\vec{k}}-\tilde{\gamma}_{-\vec{k}}\right)+p_{\vec{k}\Sigma}(\eta)\left(\chi_{-\vec{k}}-\tilde{\chi}_{-\vec{k}}\right)\right\},
\end{align}
\end{subequations}
where 
\begin{equation}
\label{eq:quadkernel}
\hat{Q} = \left(\begin{array}{cc}A_k(\eta) & E_k(\eta) \\E_k(\eta) & B_k(\eta)\end{array}\right),
\end{equation}
and we have defined the various field modes as:
\begin{equation}
\label{eq:fieldmodes}
\gamma(\vec{x}, \eta) = \int \frac{d^3 k}{(2\pi)^3}\ \gamma_{\vec{k}}(\eta) e^{-i \vec{k}\cdot\vec{x}},
\end{equation}
and similarly for $\chi(\vec{x}, \eta)$.

In terms of these modes eq.\eqref{eq:timedervlfucts} becomes
\begin{equation}
\label{eq:timedervmodes}
\gamma^{\prime}_{\vec{k}}(\eta) =-(2\pi)^3 \delta^{(3)}(\vec{k}) \phi^{\prime}(\eta), 
\end{equation}
together with the analogous equation for $\chi^{\prime}_{\vec{k}}(\eta)$.

It's worth spending a moment on a discussion of what this matrix element really is. What we have done is to take the density matrix $\rho(\eta)$ which gives $\phi,\sigma$ as the expectation values of $\Phi, \Sigma$ respectively and rewrite it in terms of the fluctuation fields $\gamma, \chi$. It is this density matrix for which $\langle \gamma(\vec{x}, \eta)\rangle= \langle \chi(\vec{x}, \eta)\rangle=0$ which is enforced by the lack of terms in the exponential proportional to $\chi_{-\vec{k}}+\tilde{\chi}_{-\vec{k}}$ and $\gamma_{-\vec{k}}+\tilde{\gamma}_{-\vec{k}}$. It can also be shown that
\begin{eqnarray}
\label{eq:littleps}
p_{\vec{k}\Phi} = (2\pi)^3 \delta^{(3)}(\vec{k}) \pi_{\Phi}=(2\pi)^3 \delta^{(3)}(\vec{k}) C(\eta) \phi^{\prime}\nonumber\\
p_{\vec{k}\Sigma } = (2\pi)^3 \delta^{(3)}(\vec{k}) \pi_{\Sigma}=(2\pi)^3 \delta^{(3)}(\vec{k}) C(\eta) \sigma^{\prime},
\end{eqnarray}
where $\pi_{\Phi}$ was defined in eq.\eqref{eq:eqsom:momenta} with a completely analogous definition for $\pi_{\Sigma}$.

The state described by eqs.\eqref{eq:densmatmatrixelts},\eqref{eq:kernels} is the density matrix version of the ABH states in \cite{Albrecht:2014aga}.

\subsection{The Liouville Equation}
\label{sec:SpicABH:liouville}

In the quadratic approximation treated here, the dynamics of this system are fully specified by the behaviors of the kernels in the matrix $\hat{Q}$ in eq.\eqref{eq:quadkernel} which in turn is obtained by solving the Liouville equation together with the zero mode equations. 

To construct the Hamiltonian in terms of the momentum modes, we start with eq.\eqref{eq:momHam:hamiltonian}, use the chain rule to rewrite $\Pi_{\Phi}$ as $\Pi_{\gamma}$ and $\Pi_{\Sigma}$ as $\Pi_{\chi}$ and then decompose these momenta into modes as
\begin{equation}
\label{eq:mommodes}
\Pi_{\gamma}=\int \frac{d^3 k}{(2\pi)^3}\ \pi_{{\vec{k}}\gamma }(\eta) e^{-i \vec{k}\cdot\vec{x}},\quad \Pi_{\chi}=\int \frac{d^3 k}{(2\pi)^3}\ \pi_{{\vec{k}}\chi}(\eta) e^{-i \vec{k}\cdot\vec{x}}.
\end{equation}
We then take the resulting Hamiltonian, expand the relevant fields in it to quadratic order in $\gamma$ and $\chi$ and rewrite it in terms of the modes $\gamma_{\vec{k}}, \chi_{\vec{k}}$. This yields:
\begin{equation}
\label{eq:quadham}
H =\int \frac{d^3 k}{(2\pi)^3}\ H_{\vec{k}}
\end{equation}
where the mode Hamiltonians $H_{\vec{k}}$ are given by
\begin{multline}
\label{eq:quadmodeham}
 H_{\vec{k}}=\frac{1}{2 C(\eta)}\left(\pi_{\vec{k}\gamma}\pi_{-\vec{k}\gamma}+\pi_{\vec{k}\chi}\pi_{-\vec{k}\chi}\right)+(2\pi)^3 \delta^{(3)}(\vec{k})\ C^2(\eta)\ \left(\partial_{\phi}V(\phi, \sigma)\gamma_{\vec{k}}+\partial_{\sigma}V(\phi, \sigma)\chi_{\vec{k}}\right)+\\
+\frac{1}{2} C(\eta) \left(\omega_{\Phi k}^2\ \gamma_{\vec{k}}\gamma_{-\vec{k}}+\omega_{\Sigma k}^2\ \chi_{\vec{k}}\chi_{-\vec{k}}\right)+\frac{1}{2}\ C^2(\eta)\ \partial_{\phi}\partial_{\sigma}V(\phi, \sigma)\left(\gamma_{\vec{k}} \chi_{-\vec{k}}+\gamma_{-\vec{k}} \chi_{\vec{k}}\right).
\end{multline}
with 
\begin{equation}
\label{eq:freqs}
\omega_{\Phi k}^2=k^2+C(\eta) \partial^2_{\phi}V(\phi,\sigma),\ \omega_{\Sigma k}^2=k^2+C(\eta) \partial^2_{\sigma}V(\phi,\sigma).
\end{equation}
Note that we have discarded the term proportional to $V(\phi, \sigma)$ as it does not contribute to the commutator in the Liouville equation.

To work in the Schr\"odinger picture, we need to know how the mode momenta act on the density matrix in eq.\eqref{eq:densmatmatrixelts}. The canonical commutation relation $\left[\Phi(\vec{x}), \Pi_{\Phi}(\vec{y})\right]=i\delta^{(3)}(\vec{x}-\vec{y})$ implies that $\pi_{\vec{k}\gamma}$ is represented on wavefunctionals as the operator:
\begin{equation}
\label{eq:momrep}
\pi_{\vec{k}\gamma}=-i(2\pi)^3 \frac{\delta}{\delta \gamma_{-\vec{k}}}
\end{equation}

Our strategy now is to use the density matrix elements in eq.\eqref{eq:densmatmatrixelts} in the calculation of both sides of the Liouville equation and then match terms that are zeroth, first and second order in the fluctuations. Note that in the calculation of $i\partial_{\eta} \langle \Phi, \Sigma \left | \rho(\eta) \right | \tilde{\Phi}, \tilde{\Sigma}\rangle$, we will need to make use of eq.\eqref{eq:timedervmodes}. Following this procedure, we arrive at the following equations:
\begin{subequations}\label{eq:Liouville}
\begin{align}
\label{eq:Liouville:zerothorder}
&\frac{N^{\prime}(\eta)}{N(\eta)}=\Omega\ \int \frac{d^3 k}{(2\pi)^3}\ \left(\frac{A_{kI}(\eta)}{C(\eta)}+\frac{B_{kI}(\eta)}{C(\eta)}\right),
\\
\label{eq:Liouville:firstorder}
&p^{\prime}_{\vec{k}\Phi}(\eta)=-(2\pi)^3 \delta^{(3)}(\vec{k})\ C^2(\eta)\ \partial_{\phi}V(\phi,\sigma),
\\
&p^{\prime}_{\vec{k}\Sigma}(\eta)=-(2\pi)^3 \delta^{(3)}(\vec{k})\ C^2(\eta)\ \partial_{\sigma}V(\phi,\sigma)
\\
\label{eq:Liouville:secondordergamma}
&i A_k^{\prime}(\eta) = \frac{A_k^2(\eta)+E_k^2(\eta)}{C(\eta)}-C(\eta)\ \omega_{k\Phi}^2,
\\
\label{eq:Liouville:secondorderchi}
&i B_k^{\prime}(\eta) = \frac{B_k^2(\eta)+E_k^2(\eta)}{C(\eta)}-C(\eta)\ \omega_{k\Sigma}^2,
\\
\label{eq:Liouville:secondordermixed}
&i E_k^{\prime}(\eta) = \frac{\left(A_k(\eta)+B_k(\eta)\right)}{C(\eta)}\ E_k(\eta)-C^2(\eta)\ \partial_{\phi}\partial_{\sigma}V(\phi,\sigma).
\end{align}
\end{subequations}
In eq.\eqref{eq:Liouville:zerothorder}, $\Omega$ denotes the comoving spatial volume, if the system were quantized in a box, and the subscript $I$ in eq.\eqref{eq:Liouville:zerothorder} denotes the imaginary part of the kernels.

Our next task is to solve the set of equations in eqs.\eqref{eq:Liouville} and understand how entanglement might arise.

\subsection{Mode Equations and the Origin of Entanglement}
\label{sec:SpicABH:modeeqs}

We are used to computing inflationary $n$-point functions in terms of the BD modes. However, in the Schr\"odinger picture we are using, these correlation functions would be written in terms of the kernels $A_k(\eta), B_k(\eta), E_k(\eta)$. How do we see the BD modes arising from this formalism? The key is in noticing that the equations for $A_k(\eta), B_k(\eta)$ are of Riccati type and hence can be transformed from a first order non-linear equation to a second order {\em linear} equation via the definitions:
\begin{eqnarray}
\label{eq:modedefs}
i A_k(\eta) &\equiv& C(\eta)\left(\frac{f^{\prime}_k(\eta)}{f_k(\eta)}-\frac{1}{2} \frac{C^{\prime}(\eta)}{C(\eta)}\right)\nonumber\\
i B_k(\eta) &\equiv& C(\eta)\left(\frac{g^{\prime}_k(\eta)}{g_k(\eta)}-\frac{1}{2} \frac{C^{\prime}(\eta)}{C(\eta)}\right).
\end{eqnarray}
In terms of $f_k, g_k$ the kernel equations become
\begin{eqnarray}
\label{eq:modeeqs}
&&\frac{f^{\prime \prime}(\eta)}{f_k(\eta)}+\left( \omega_{k\Phi}^2-\frac{1}{6} C(\eta)R\right)=\frac{E_k^2(\eta)}{C^2(\eta)}\nonumber\\
&&\frac{g^{\prime \prime}(\eta)}{g_k(\eta)}+\left( \omega_{k\Sigma}^2-\frac{1}{6} C(\eta)R\right)=\frac{E_k^2(\eta)}{C^2(\eta)}
\end{eqnarray}
where $R$ is the curvature scalar $C(\eta) R = 3 \left(D^{\prime}(\eta)+1\slash 2\ D^2(\eta)\right)$, while the entanglement kernel satisfies
\begin{equation}
\label{eq:entker}
E_k^{\prime}(\eta) = -\left(\frac{f^{\prime}_k(\eta)}{f_k(\eta)}+\frac{g^{\prime}_k(\eta)}{g_k(\eta)}-\frac{C^{\prime}(\eta)}{C(\eta)}\right)\ E_k(\eta)+i C^2(\eta)\ \partial_{\phi}\partial_{\sigma}V(\phi,\sigma).
\end{equation}

Eq.\eqref{eq:entker} admits an integrating factor to become
\begin{equation}
\label{eq:entkerint}
\partial_{\eta}\left(\frac{f_k(\eta) g_k(\eta)}{C(\eta)} E_k(\eta)\right)=\frac{f_k(\eta) g_k(\eta)}{C(\eta)} \left(i C^2(\eta)\ \partial_{\phi}\partial_{\sigma}V(\phi,\sigma)\right),
\end{equation}
which can be integrated to find
\begin{equation}
\label{eq:entklerfullint}
\frac{f_k(\eta) g_k(\eta)}{C(\eta)} E_k(\eta)=\lambda_k + i\int^{\eta}_{\eta_0}d\eta^{\prime}\ \left[f_k(\eta^{\prime})\  g_k(\eta^{\prime})\ C(\eta^{\prime})\ \partial_{\phi}\partial_{\sigma}V(\phi(\eta^{\prime}),\sigma(\eta^{\prime}))\right],
\end{equation}
where $\eta_0$ is the initial time we start the system at. 

In previous work, only the so-called entanglement parameter $\lambda_k$ appeared. It was a free parameter that could be constrained by requirements such as the energy density coming from entanglement being small enough to not overwhelm the inflationary energy density\cite{Albrecht:2014aga}. Now, however, we see that even if we set the initial entanglement to zero, so that $\lambda_k$ vanishes, we can {\em still} generate a non-zero entanglement kernel as long as the mixed partial $\partial_{\phi}\partial_{\sigma}V(\phi(\eta^{\prime}),\sigma(\eta^{\prime}))$ is non-vanishing along the trajectory defined by the equations of motion for the expectation values $\phi, \sigma$ in eq.\eqref{eq:quadzeromodeeom}. 

It's worth considering some examples here. First, consider just a mass mixing term of the form $\mu^2 \phi\ \sigma$ together with mass terms for $\phi$ and $\sigma$ separately. Then the mixed partial is just $\mu^2$ and entanglement gets generated. What we see here is just the fact that $\phi$ and $\sigma$ are not mass eigenstates and the entanglement here just comes from the fact that we want to measure the correlation functions of the non-mass eigenstates (for example, they might be what generates metric perturbations). This is fully analogous to what happens in neutrino oscillations.

Next take the case of hybrid inflation as in ref.\cite{Linde:1993cn}. The potential is given by
\begin{equation}
\label{eq:potentialhybrid}
V(\phi,\sigma) = \frac{1}{4\lambda}\left(M^2-\lambda\sigma^2\right)^2+\frac{1}{2} m^2 \phi^2 + \frac{g^2}{2}\phi^2 \sigma^2.
\end{equation}
The trajectory taken during inflation is $\sigma=0$ with $\phi$ rolling. It is easy to see that the mixed partial {\em vanishes} along this trajectory so that entanglement {\em cannot} be generated, at least not via the mechanism described above, in this system. 

In order to obtain a quantitative picture of the evolution of the system, the correlation functions such as $\langle \gamma^2\rangle,\langle \chi^2\rangle, \langle \gamma \chi\rangle$ need to be computed. These are integrals of various combinations of the kernels $A_k,B_k, E_k$ as shown in ref.\cite{Albrecht:2014aga}, but now themselves depend on the evolution of the expectation values. Thus, even within the quadratic approximation this systems exhibits a good deal of complexity. We defer the numerical analysis of this system to later work\cite{RHBRzeromodes}.

\section{Entangling $\zeta$}
\label{sec:entzeta}
How does the above analysis change if one of the fields is the metric perturbation $\zeta$? We might, for example, like to be able to dynamically generate entanglement between $\zeta$ and a scalar $\Sigma$ so as to use it to introduce features into the power spectrum or the bispectrum. However, we do not expect $\zeta$ to have an expectation value (in fact, $\zeta$ is defined so that its tadpole vanishes) so the above method does not apply, at least not directly.

\subsection{$\zeta$-Scalar Hamiltonian at Cubic Order and the Liouville Equation}
\label{subsec:cubicham}

The key to generating entanglement in the two scalar case was to consider the interactions between the two scalar fields; mimicking this for $\zeta$ requires constructing the interaction Lagrangian between $\zeta$ and $\Sigma$. Thankfully, this has already been done to cubic order in refs.\cite{Weinberg:2005vy,delRio:2018vrj}:
\begin{equation}
\label{eq:cubiczetascalar}
S^{(3)}_{\zeta\Sigma} = \int d^4 x\ C(\eta)\epsilon\left\{\frac{\zeta}{2}\left(\Sigma^{\prime 2} +\partial_i \Sigma\ \partial_i \Sigma\right)-\Sigma^{\prime}\ \partial_i \Sigma\ \partial_i \partial^{-2}\zeta^{\prime}\right\}.
\end{equation}
We will also need the quadratic $\zeta$ action so as to compute the momentum conjugate to $\zeta$:
\begin{equation}
\label{eq:zetaquadaction}
S^{(2)}_{\zeta} =  M^2_{\rm Pl}\int d^4 x\  C(\eta)\epsilon\left(\zeta^{\prime 2}-(\partial \zeta)^2\right),
\end{equation}
where $\epsilon$ is the usual inflationary slow-roll parameter. This last equation tells us that
\begin{equation}
\label{eq:zetamom}
\Pi_{\zeta}=2 M^2_{\rm Pl}\  C(\eta)\epsilon\ \zeta^{\prime}. 
\end{equation}
As in the two scalar case, we take $\langle \Sigma \rangle(\eta)=\sigma(\eta)$ and consider the parts of $S^{(3)}_{\zeta\Sigma}$ that can generate a cross term quadratic in the fluctuations. Out of the three terms only the first and last do so since the middle term requires two spatial derivatives of the scalar field which implies that we cannot replace $\Sigma$ by $\sigma(\eta)$ to generate a quadratic term.

Constructing the quadratic approximation to the Hamiltonian for $\zeta$ coupled to the fluctuations $\chi$ of $\Sigma$ around $\sigma$ from these actions yields:
\begin{equation}
H =H^{(1)}+ H^{(2)}+H^{(3)},
\end{equation}
with
\begin{subequations}\label{eq:Zetascalarham}
\begin{align}
\label{eq:Zetascalarham:linear}
&H^{(1)} =-\sigma^{\prime 2}(\eta)\ C(\eta)\ \epsilon \int \frac{d^3 k}{(2\pi)^3}\ \zeta_{\vec{k}},
\\
\label{eq:Zetascalarham:quad}
&H^{(2)}=\int \frac{d^3 k}{(2\pi)^3}\ \left\{\frac{\pi_{\vec{k}\zeta}\pi_{-\vec{k}\zeta}}{4 C(\eta)M^2_{\rm Pl}\epsilon}+ C(\eta)M^2_{\rm Pl}\ \epsilon\ k^2 \zeta_{\vec{k}}\zeta_{-\vec{k}}+\right. \nonumber\\
&\left . +\frac{\pi_{\vec{k}\chi}\pi_{-\vec{k}\chi}}{2 C(\eta)}+(2\pi)^3 \delta^{(3)}(\vec{k})\ C^2(\eta)\ \partial_{\sigma}V( \sigma)\chi_{\vec{k}}+\frac{1}{2} C(\eta)\ \omega_{k\Sigma }^2\ \chi_{\vec{k}}\chi_{-\vec{k}}\right\},
\\
\label{eq:Zetascalarham:cubic}
&H^{(3)}=-\int \frac{d^3 k}{(2\pi)^3}\ \sigma^{\prime}(\eta)\left\{\epsilon\ \zeta_{\vec{k}}\ \pi_{-\vec{k}\chi}+\frac{1}{2 M^2_{\rm Pl}}\chi_{\vec{k}} \ \pi_{-\vec{k}\zeta}\right\}.
\end{align}
\end{subequations}
We can reduce the amount of work required in computing both sides of the Liouville equation by redefining the $\zeta_{\vec{k}}, \pi_{\vec{k} \zeta}$ to absorb the factors of $\epsilon$ and $M^2_{\rm Pl}$ appearing in eq.
\eqref{eq:Zetascalarham:quad}. Set $\tilde{\zeta}_{\vec{k}}= \sqrt{2 \epsilon M^2_{\rm Pl}}\ \zeta_{\vec{k}}$ and $\tilde{\pi}_{\vec{k} \zeta} = \pi_{\vec{k} \zeta}\slash \sqrt{2 \epsilon M^2_{\rm Pl}}$. This preserves the canonical commutation relations and transforms $H^{(2)}$ into a form that is identical to that in eq.\eqref{eq:quadmodeham} with the replacements $\gamma\leftrightarrow \tilde{\zeta},\pi_{\gamma}\leftrightarrow \tilde{\pi}_{\zeta}$ as well as taking $V(\phi,\sigma)=V(\sigma)$. In terms of the tilded variables, $H^{(3)}$ becomes
\begin{equation}
H^{(3)}=-\sqrt{\frac{\epsilon}{2 M^2_{\rm Pl}}}\sigma^{\prime}(\eta) \int \frac{d^3 k}{(2\pi)^3}\ \left\{\tilde{\zeta}_{\vec{k}} \pi_{-\vec{k}\chi}+\chi_{\vec{k}} \tilde{\pi}_{-\vec{k} \zeta}\right\}.
\end{equation}
We can write the entangled state just as in eqs.\eqref{eq:densmatmatrixelts}-\eqref{eq:kernels}, again with the replacements above. We write the $A_k,E_k$ kernels as $\tilde{A}_k,\tilde{E}_k$ to remind ourselves that they are the kernels corresponding to the tilded variables; this becomes an issue (though clearly not a major one!) because the correlation functions we usually need are those in terms of the original $\zeta$ not $\tilde{\zeta}$. 

We now follow the same procedure as we did for the two scalar case: compute both sides of the Liouville equation and match powers of fluctuation modes. Doing this yields equations similar to eqs.\eqref{eq:Liouville}; the ones most relevant to the current discussion are:
\begin{subequations}\label{eq:Liouvillezeta}
\begin{align}
\label{eq:Liouville:secondorderzeta}
&i \tilde{A}_k^{\prime}(\eta) = \frac{\tilde{A}_k^2(\eta)+\tilde{E}_k^2(\eta)}{C(\eta)}-C(\eta)\ \omega_{k\zeta}^2,
\\
\label{eq:Liouville:secondorderchi}
&i B_k^{\prime}(\eta) = \frac{B_k^2(\eta)+\tilde{E}_k^2(\eta)}{C(\eta)}-C(\eta)\ \omega_{k\Sigma}^2,
\\
\label{eq:Liouville:secondordermixed}
&i \tilde{E}_k^{\prime}(\eta) = \frac{\left(\tilde{A}_k(\eta)+B_k(\eta)\right)}{C(\eta)}\ \tilde{E}_k(\eta)+i \sqrt{\frac{\epsilon}{2 M^2_{\rm Pl}}} \sigma^{\prime}(\eta)\left(\tilde{A}_k(\eta)+B_k(\eta)\right) .
\end{align}
\end{subequations}
Once again we have Riccati equations and can define the standard modes $\tilde{f}_k, g_k$ analogously to what was done in eqs.\eqref{eq:modedefs} to turn these equations into second order ones. The equation for the entanglement kernel $\tilde{E}_k$ again admits and integrating factor and we find that (assuming a slowly varying $\epsilon$)
\begin{equation}
\label{eq:tildeE}
\frac{\tilde{f}_k(\eta) g_k(\eta)}{C(\eta)} \tilde{E}_k(\eta)=\lambda_k +\sqrt{\frac{\epsilon}{2 M^2_{\rm Pl}}}\ \int^{\eta}_{\eta_0}d\eta^{\prime}\ \sigma^{\prime}(\eta^{\prime})\ \partial_{\eta^{\prime}}\ln\left(\frac{\tilde{f}_k(\eta^{\prime})\  g_k(\eta^{\prime})}{C(\eta^{\prime})}\right),
\end{equation}
The quadratic Hamiltonian obtained from the cubic interaction between $\zeta$ and a scalar induces a source term that, just as in the two-scalar case, can generate a non-zero entanglement starting from a vanishing initial value. 

\subsection{Entangling $\zeta$ and tensor fluctuations}

In refs.\cite{Bolis:2016vas,Collins:2016ahj} entanglement between $\zeta$ and the tensor fluctuations $h_{i j}$ was considered, in an attempt to explain anomalies in CMB correlations. Is there a way to generate such an entangling term in the state? One possibility is to go to quartic order in the fluctuations between $\zeta$, $h_{i j}$ and a scalar field $\Sigma$. The lowest order term that appears in the action is of the form\cite{delRio:2018vrj}:
\begin{equation}
\label{eq:quartic}
S^{(4)}_{\zeta h \Sigma} = -\frac {1}{2}\int d^4 x\ \epsilon  C(\eta) \zeta \left(h_{i j}\partial_i \Sigma \partial_j \Sigma\right).
\end{equation}
To use this to generate the quadratic term we want, i.e. the one that couples $\zeta$ to $h_{i j}$ would require a spatially varying background. This may be of interest from a number of perspectives, but we defer the exploration of this possibility to later work\cite{RHBRzeromodes}.

\section{Conclusions}
\label{sec:conclusions}
The inflationary paradigm has a great deal of explanatory power. It gives us a coherent story of how quantum fluctuations become metric perturbations that is both qualitatively {\em and} quantitatively consistent with observations. However, a linchpin of this idea is that we can essentially use the principle of equivalence to say that since at short distances and times spacetime should be locally flat, we should therefore only consider quantum states whose asymptotic behavior approaches that of the flat space vacuum state. This is the BD prescription. 

We would argue that this is too narrow a view of the situation. After all, inflation presumably begins at some finite time in the past and it may or may not be reasonable to apply the BD prescription in this case. Moreover, if we want to persuade ourselves that the BD state {\em is} the one we should use, we need to argue that all others are inconsistent, either internally or with available data. Thus, an exploration of the type we have embarked on, where we consider the physics of a variety of non-BD states, is not unreasonable. In particular, this serves as motivation for studying states such as the ABH ones.

One fly in the ointment concerning the ABH states had to do with how the entanglement was generated. It was all well and good to view them as possible allowed states in the space of effective inflationary states, but this did not obviate the need for an explanation of the origin of these states. These issues are at least partially resolved in this work, where we show that interactions can in fact generate entanglements. We were able to argue within the Gaussian approximation that, if the fields had rolling expectation values, then the quadratic fluctuation Hamiltonian generated by expanding around these expectation values gave rise to source terms for the entanglement. As a by-product, we were able to generalize the procedure for computing the time evolution of these states in the presence of rolling expectation values. 

There are a number of interesting directions this work can be taken in. One is to follow up on the discussion of how $\zeta$ and tensor fluctuations might wind up entangled and consider the possibility of spatially varying scalar field backgrounds. Another is to allow for the possibility of entangling $\zeta$ with a mixed state such as a thermal one. Work on this is in progress\cite{RHBRmixedentanglement}.

\bibliographystyle{JHEP}
\bibliography{entangledslowroll}

\providecommand{\href}[2]{#2}\begingroup\raggedright\begin{thebibliography}{10}

\bibitem{Ade:2015lrj}
{\scshape Planck} collaboration, P.~A.~R. Ade et~al., \emph{{Planck 2015
  results. XX. Constraints on inflation}},
  \href{https://arxiv.org/abs/1502.02114}{{\ttfamily 1502.02114}}.

\bibitem{Akrami:2018odb}
{\scshape Planck} collaboration, Y.~Akrami et~al., \emph{{Planck 2018 results.
  X. Constraints on inflation}},
  \href{https://arxiv.org/abs/1807.06211}{{\ttfamily 1807.06211}}.

\bibitem{Ade:2015hxq}
{\scshape Planck} collaboration, P.~A.~R. Ade et~al., \emph{{Planck 2015
  results. XVI. Isotropy and statistics of the CMB}},
  \href{https://arxiv.org/abs/1506.07135}{{\ttfamily 1506.07135}}.

\bibitem{Bunch:1978yq}
T.~S. Bunch and P.~C.~W. Davies, \emph{{Quantum Field Theory in de Sitter
  Space: Renormalization by Point Splitting}},
  \href{https://doi.org/10.1098/rspa.1978.0060}{\emph{Proc. Roy. Soc. Lond.}
  {\bfseries A360} (1978) 117}.

\bibitem{Birrell:1982ix}
N.~D. Birrell and P.~C.~W. Davies, \emph{{Quantum Fields in Curved Space}},
  Cambridge Monographs on Mathematical Physics. Cambridge Univ. Press,
  Cambridge, UK, 1984,
  \href{https://doi.org/10.1017/CBO9780511622632}{10.1017/CBO9780511622632}.

\bibitem{Albrecht:2018hoh}
A.~Albrecht, N.~Bolis and R.~Holman, \emph{{Cosmic Inflation: The Most Powerful
  Microscope in the Universe}},
  \href{https://arxiv.org/abs/1806.00392}{{\ttfamily 1806.00392}}.

\bibitem{Kaloper:2018zgi}
N.~Kaloper and J.~Scargill, \emph{{Quantum Cosmic No-Hair Theorem and
  Inflation}},  \href{https://arxiv.org/abs/1802.09554}{{\ttfamily
  1802.09554}}.

\bibitem{Albrecht:2014aga}
A.~Albrecht, N.~Bolis and R.~Holman, \emph{{Cosmological Consequences of
  Initial State Entanglement}},
  \href{https://doi.org/10.1007/JHEP11(2014)093}{\emph{JHEP} {\bfseries 11}
  (2014) 093} [\href{https://arxiv.org/abs/1408.6859}{{\ttfamily 1408.6859}}].

\bibitem{usbispectrum}
A.~Albrecht, N.~Bolis and R.~Holman, \emph{Entangled bi-spectra, (in
  progress)}, .

\bibitem{Bolis:2016vas}
N.~Bolis, A.~Albrecht and R.~Holman, \emph{{Modifications to Cosmological Power
  Spectra from Scalar-Tensor Entanglement and their Observational
  Consequences}}, \href{https://doi.org/10.1088/1475-7516/2017/08/E01,
  10.1088/1475-7516/2016/12/011}{\emph{JCAP} {\bfseries 1612} (2016) 011}
  [\href{https://arxiv.org/abs/1605.01008}{{\ttfamily 1605.01008}}].

\bibitem{Collins:2016ahj}
H.~Collins and T.~Vardanyan, \emph{{Entangled Scalar and Tensor Fluctuations
  during Inflation}},
  \href{https://doi.org/10.1088/1475-7516/2016/11/059}{\emph{JCAP} {\bfseries
  1611} (2016) 059} [\href{https://arxiv.org/abs/1601.05415}{{\ttfamily
  1601.05415}}].

\bibitem{Bolis:2018jmo}
N.~Bolis, T.~Fujita, S.~Mizuno and S.~Mukohyama, \emph{{Quantum Entanglement in
  Multi-field Inflation}},
  \href{https://doi.org/10.1088/1475-7516/2018/09/004}{\emph{JCAP} {\bfseries
  1809} (2018) 004} [\href{https://arxiv.org/abs/1805.09448}{{\ttfamily
  1805.09448}}].

\bibitem{Phillips:2014yma}
D.~Phillips, A.~Scacco and A.~Albrecht, \emph{{Holographic bounds and finite
  inflation}}, \href{https://doi.org/10.1103/PhysRevD.91.043513}{\emph{Phys.
  Rev.} {\bfseries D91} (2015) 043513}
  [\href{https://arxiv.org/abs/1410.6065}{{\ttfamily 1410.6065}}].

\bibitem{Kanno:2014ifa}
S.~Kanno, \emph{{Impact of quantum entanglement on spectrum of cosmological
  fluctuations}},
  \href{https://doi.org/10.1088/1475-7516/2014/07/029}{\emph{JCAP} {\bfseries
  1407} (2014) 029} [\href{https://arxiv.org/abs/1405.7793}{{\ttfamily
  1405.7793}}].

\bibitem{Boyanovsky:1993xf}
D.~Boyanovsky, H.~J. de~Vega and R.~Holman, \emph{{Nonequilibrium evolution of
  scalar fields in FRW cosmologies I}},
  \href{https://doi.org/10.1103/PhysRevD.49.2769}{\emph{Phys. Rev.} {\bfseries
  D49} (1994) 2769} [\href{https://arxiv.org/abs/hep-ph/9310319}{{\ttfamily
  hep-ph/9310319}}].

\bibitem{RHBRmixedentanglement}
R.~Holman and B.~J. Richard, \emph{Mixed entangled states (in progress)}, .

\bibitem{Linde:1993cn}
A.~D. Linde, \emph{{Hybrid inflation}},
  \href{https://doi.org/10.1103/PhysRevD.49.748}{\emph{Phys. Rev.} {\bfseries
  D49} (1994) 748} [\href{https://arxiv.org/abs/astro-ph/9307002}{{\ttfamily
  astro-ph/9307002}}].

\bibitem{RHBRzeromodes}
R.~Holman and B.~J. Richard, \emph{Entangled slow roll (in progress)}, .

\bibitem{Weinberg:2005vy}
S.~Weinberg, \emph{{Quantum contributions to cosmological correlations}},
  \href{https://doi.org/10.1103/PhysRevD.72.043514}{\emph{Phys. Rev.}
  {\bfseries D72} (2005) 043514}
  [\href{https://arxiv.org/abs/hep-th/0506236}{{\ttfamily hep-th/0506236}}].

\bibitem{delRio:2018vrj}
A.~del Rio, R.~Durrer and S.~P. Patil, \emph{{Tensor Bounds on the Hidden
  Universe}}, \href{https://doi.org/10.1007/JHEP12(2018)094}{\emph{JHEP}
  {\bfseries 12} (2018) 094}
  [\href{https://arxiv.org/abs/1808.09282}{{\ttfamily 1808.09282}}].

\end{thebibliography}\endgroup

\end{document}